\documentstyle[epsfig]{article}
\title{Schr\"odinger-Wheeler-DeWitt equation in chaplygin gas FRW cosmological model}
\author{P. Pedram, S. Jalalzadeh\thanks{Email: s-jalalzadeh@sbu.ac.ir}\,\, and S. S. Gousheh
\\ {\small Department of Physics, Shahid Beheshti University,
Evin, Tehran 19839, Iran}}
\begin{document}
\maketitle \baselineskip 24pt
\begin{abstract}
We present a chaplygin gas Friedmann-Robertson-Walker quantum
cosmological model. In this work the Schutz's variational formalism
is applied with positive, negative, and zero constant spatial
curvature. In this approach the notion of time can be recovered.
These give rise to Schr\"odinger-Wheeler-DeWitt equation for the
scale factor. We use the eigenfunctions in order to construct wave
packets for each case. We study the time dependent behavior of the
expectation value of the scale factor, using the many-worlds
interpretations of quantum mechanics.
\end{abstract}

\textit{Pacs}:{98.80.Qc, 04.40.Nr, 04.60.Ds}

\section{Introduction}
In recent years supernova Ia  (SNIa) observations show that the expansion of the universe is
accelerating \cite{Riess:1998cb} contrary to Friedmann-Robertson-Walker (FRW) cosmological models,
with non-relativistic matter and radiation. Also cosmic microwave background radiation (CMBR) data
\cite{Spergel:2003cb,2a} is suggesting  that the expansion of our universe seems to be in an
accelerated state. This is referred to ``dark energy'' effect \cite{3a}. Cosmological constant,
$\Lambda $, as usual vacuum energy can be responsible for this evolution by providing a negative
pressure \cite{3b,3c}. Unfortunately, the observed value of $\Lambda$ is $120$ orders of magnitude
smaller than the one computed from field theory methods \cite{3b,3c}. Quintessence is an
alternative to consider a dynamical vacuum energy \cite{Wetterich:fm}, involving one or two scalar
fields, some with potentials justified from supergravity theories \cite{Brax:1999yv}. However, the
fine-tuning problem of these models which arise from cosmic coincidence issue has no satisfactory
solution.

The Chaplygin gas model is an interesting proposal \cite{Kamenshchik}, describing a transition from a universe filled with dust-like matter to an
accelerating expanding stage. This model was later generalized in Ref.~\cite{Bilic,A}. The generalized Chaplygin gas model is described by a perfect
fluid obeying an exotic equation of state \cite{A}
\begin{equation}
p=-\frac{A}{\rho ^{\alpha }},  \label{cgi1}
\end{equation}
where $A$ is a positive constant and $0<\alpha \leq 1$. The standard
Chaplygin gas \cite{Kamenshchik} corresponds to $\alpha =1$. Some
publications
\cite{Bento,3rp,2rp,Fabris,C,Ogawa,NewBD,18a,20a,21a,23a,27a,Rev1,Rev2,DySy,Jackiw,setare1,setare2,setare3}
and reviews \cite{Rev1,Rev2} which studied the Chaplygin gas
cosmological models have already appeared in the literature.

Recently, quantum mechanical description of a FRW model with a generalized Chaplygin gas has been discussed in Ref. \cite{Buahmadi} in order to
retrieve explicit mathematical expressions for the different quantum mechanical states and determine the transition probabilities towards an
accelerated stage. In this paper we investigate the existence of singularities at quantum level in Chaplygin gas cosmological models. In the quantum
cosmology the Wheeler-DeWitt (WD) equation in minisuperspace which determines the wave function of the Universe, can be constructed using ADM
decomposition of the geometry \cite{7} in the Hamiltonian formalism of general relativity.

The presence of matter in quantum cosmology needs further consideration and can be described by
fundamental fields, as done in Ref. \cite{10}. Using WKB approximation one can predict the behavior
of the quantum universe which leads to determination of the trajectories in phase space. However,
even in the minisuperspace, general exact solutions are hard to find, the Hilbert space structure
is obscure and it is a subtle matter to recover the notion of a semiclassical time \cite{8,10}.

In the present work, we describe matter as a chaplygin gas. This description is essentially
semiclassical from the start, but it has the advantage of furnishing a variable, connected with the
matter degrees of freedom, which can naturally be identified with time, leading to a well-defined
Hilbert space structure. It is very convenient to construct a quantum chaplygin gas model. Schutz's
formalism \cite{11,12} gives dynamics to the fluid degrees of freedom in interaction with the
gravitational field. Using proper canonical transformations, at least one conjugate momentum
operator associated with matter appears linearly in the action integral. Therefore, a
Schr\"odinger-like equation can be obtained with the matter variable playing the role of time.

Here, we use the formalism of quantum cosmology in order to quantize three
Friedmann-Robertson-Walker chaplygin gas models in the presence of a negative cosmological
constant. In Sec. 2 the quantum cosmological model with a chaplygin gas as the matter content is
constructed in Schutz's formalism \cite{20}, and the Schr\"odinger-Wheeler-DeWitt (SWD) equation in
minisuperspace is written down to quantize the model. The wave-function depends on the scale factor
$a$ and on the canonical variable associated to the fluid, which in the Schutz variational
formalism plays the role of time $T$. We separate the wave-function into two parts, one depending
solely on the scale factor and the other depending only on the time. The solution in the time
sector of the SWD equation is trivial, leading to imaginary exponentials of the type $e^{-iEt}$,
where $E$ is the system energy and $t =- T$. In Sec. 4 we construct wave packets from the
eigenfunctions and compute the time-dependent expectation values of the scale factors. In Sec. 5,
we present our conclusions.

\section{Model}
We need the Hamiltonian for a chaplygin gas model in the formalism developed by Schutz. The starting point is the action for gravity plus chaplygin
gas, which in this formalism is written as
\begin{eqnarray}
\label{action} S = \int_Md^4x\sqrt{-g}\, R + 2\int_{\partial M}d^3x\sqrt{h}\, h_{ab}\, K^{ab} + \int_Md^4x\sqrt{-g}\,\, p \quad ,
\end{eqnarray}
where $K^{ab}$ is the extrinsic curvature and $h_{ab}$ is the induced metric over the
three-dimensional spatial hypersurface, which is the boundary $\partial M$ of the four dimensional
manifold $M$. Units are chosen such that the factor $16\pi G$ becomes equal to one. The first two
terms were first obtained in \cite{7}; the last term of (\ref{action}) represents the matter
contribution to the total action and $p$ is the pressure. In Schutz's formalism \cite{11,12} the
fluid's four-velocity is expressed in terms of five potentials $\Phi$, $\zeta$, $\beta$, $\theta$
and $S$
\begin{equation}
u_\nu = \frac{1}{\mu}(\Phi_{,\nu} + \zeta\beta_{,\nu} + \theta S_{,\nu})
\end{equation}
where $\mu$ is the specific enthalpy. The variable $S$ is the specific entropy, while the potentials $\zeta$ and $\beta$ are connected with rotation
and are absent in models of the Friedmann-Robertson-Walker (FRW) type. The variables $\Phi$ and $\theta$ have no clear physical meaning. The
four-velocity is subject to the normalization condition
\begin{equation}
u^\nu u_\nu = -1.
\end{equation}
The FRW metric
\begin{equation}
ds^2 = - N^2(t)dt^2 + a^2(t)g_{ij}dx^idx^j,
\end{equation}
is now inserted in the action (\ref{action}). In this expression, $N(t)$ is the lapse function and $g_{ij}$ is the metric on the constant-curvature
spatial section.

Following the thermodynamic description of Ref.~\cite{14}, the basic thermodynamic relations take the form
\begin{eqnarray}
  \rho &=& \rho_0[1+\Pi], \quad h=1+\Pi+p/\rho_0 \\ \nonumber
  \tau dS &=&
  d\Pi+p\,d(1/\rho_0)\\
  &=&\frac{(1+\Pi)^{-\alpha}}{1+\alpha}d\left[(1+\Pi)^{1+\alpha}+\frac{A}{\rho_0^{1+\alpha} }\right]
\end{eqnarray}
It then follows that to within a factor
\begin{eqnarray}
  \tau &=& \frac{(1+\Pi)^{-\alpha}}{1+\alpha} \\
  S &=& (1+\Pi)^{1+\alpha}+\frac{A}{\rho_0^{1+\alpha}}.
\end{eqnarray}
Therefore, the equation of state takes the form
\begin{equation}
   p=-A\left[\frac{1}{A}\left(1-\frac{\,\,h^{\frac{\alpha+1}{\alpha}}}{S^{1/\alpha}}\right)\right]^{\frac{\alpha+1}{\alpha}}
\end{equation}
The particle number density and energy density are, respectively,
\begin{eqnarray}
  \rho &=& \left[\frac{1}{A}\left(1-\frac{\,\,h^{\frac{\alpha+1}{\alpha}}}{S^{1/\alpha}} \right) \right]^{\frac{-1}{\,\,1+\alpha}} \\
  \rho_0 &=& \frac{\rho+p}{h}
\end{eqnarray}
where $h=(\dot{\Phi}+\theta\dot{S})/N$. After dropping the surface terms, the final reduced action takes the form
\begin{eqnarray}
S = \int dt\biggr\{-6\frac{\dot a^2a}{N} + 6kNa -N a^3
A\left[\frac{1}{A}\left(1-\frac{\,\,(\dot{\Phi}+\theta\dot{S})^{\frac{\alpha+1}{\alpha}}}{N^{\frac{\alpha+1}{\alpha}}
S^{1/\alpha}}\right)\right]^{\frac{\alpha+1}{\alpha}}\biggr\}.
\end{eqnarray}
The reduced action may be further simplified using canonical methods \cite{14}, resulting in the super-Hamiltonian
\begin{equation}
{\cal H} = - \frac{p_a^2}{24a} -6ka +\left(S p_{\Phi}^{1+\alpha}+A a^{3(1+\alpha)} \right)^{\frac{1}{1+\alpha}}
\end{equation}
where $p_a= -12{\dot aa}/{N}$ and $p_\Phi =\frac{\partial{\cal
L}}{\partial \dot{\Phi}}\,$. However, an analytical quantum
mechanical treatment of this FRW minisuperspace with the above
Hamiltonian does not seem feasible. Therefore, it requires the
following approximation \cite{Buahmadi},
\begin{eqnarray}
\hspace{-1cm}\left(S p_{\Phi}^{1+\alpha}+ A a^{3(1+\alpha)}\right) ^{\frac1{1+\alpha}}\approx S^{\frac{1}{1+\alpha}} p_{\Phi}\left[
1+\frac1{1+\alpha}\frac{ Aa^{3(\alpha+1)}}{S p_{\Phi}^{1+\alpha}}+\right. \left.\frac12\frac1{1+\alpha}\left( \frac1{1+\alpha}-1\right)
\frac{A^{2}}{S^2 p_{\Phi}^{2(1+\alpha)}}a^{6(\alpha+1)}+\ldots\right]. \nonumber \\
\end{eqnarray}
Hence, up to the leading order, the super-Hamiltonian takes the form
\begin{equation}
{\cal H} = - \frac{p_a^2}{24a}-6ka +S^{\frac{1}{1+\alpha}} p_{\Phi}
\end{equation}
 The following additional canonical transformations,
\begin{eqnarray}
T &=& -(1+\alpha)p_\Phi^{-1}  S^{\frac{\alpha}{1+\alpha}}p_S, \quad \quad p_T =S^{\frac{1}{1+\alpha}} p_\Phi,
\end{eqnarray}
simplifies the super-Hamiltonian to,
\begin{equation}
{\cal H} = - \frac{p_a^2}{24a} -6ka + p_T , ,\label{EqHamiltonian}
\end{equation}
where the momentum $p_T$ is the only remaining canonical variable associated with matter. It appears linearly in the super-Hamiltonian. The parameter
$k$ defines the curvature of the spatial section, taking the values $0, 1, - 1$ for a flat, positive-curvature or negative-curvature Universe,
respectively.

The classical dynamics is governed by the Hamilton equations, derived from Eq. (\ref{EqHamiltonian}) and Poisson brackets, namely
\begin{equation}
\left\{
\begin{array}{llllll}
\dot{a} =&\{a,N{\cal H}\}=-\frac{\displaystyle Np_{a}}{\displaystyle 12a}\, ,\\
 & \\
\dot{p_{a}} =&\{p_{a},N{\cal H}\}=- \frac{N}{24a^2}p_a^2+6Nk  \\
& \\
\dot{T} =&\{T,N{\cal H}\}=N\, ,\\
 & \\
\dot{p_{T}} =&\{p_{T},N{\cal H}\}=0\, .\\
& \\
\end{array}
\right. \label{4}
\end{equation}
We also have the constraint equation ${\cal H} = 0$. Choosing the
gauge $N=1$, we have the following solutions for the system
\begin{eqnarray}
\ddot{a}&=&-\frac{\dot a^2}{2a}-\frac{k}{2a},\\
0&=&-6a\dot a^2-6k a+\,p_T.
\end{eqnarray}
Imposing the standard quantization conditions on the canonical
momenta and demanding that the super-Hamiltonian operator annihilate
the wave function, we are led to the following SWD equation in
minisuperspace ($\hbar =1$)
\begin{equation}
\label{sle} \frac{\partial^2\Psi}{\partial a^2} - 144ka^2\Psi + i24a\frac{\partial\Psi}{\partial t} = 0 \quad .
\end{equation}
In this equation, $t=-T$ corresponds to the time coordinate. As discussed in \cite{nivaldo,15}, in order for  the Hamiltonian operator ${\hat H}$ to
be self-adjoint  the inner product of any two wave functions $\Phi$ and $\Psi$ must take the form
\begin{equation}
(\Phi,\Psi) = \int_0^\infty a\,\Phi^*\Psi da,
\end{equation}
Moreover,  the wave functions should satisfy the restrictive boundary conditions
\begin{equation}
\label{boundary} \Psi(0,t) = 0 \quad \mbox{or} \quad \frac{\partial\Psi (a,t)}{\partial a}\bigg\vert_{a = 0} = 0.
\end{equation}
The SWD equation (\ref{sle}) can  be solved by separation of
variables as follows
\begin{equation}
\psi(a,t) = e^{-iEt}\psi(a) \label{11}
\end{equation}
where the $a$ dependent part of the wave function ($\psi(a)$) satisfies
\begin{equation}
\label{sle2} -\psi''(a) +144 ka^2\psi(a) =24Ea\,\psi(a),
\end{equation}
and the prime means derivative with respect to $a$.

\section{Results}
For $k = 0$ the time-independent  Wheeler-DeWitt equation (\ref{sle2}) reduces to
\begin{equation}
\label{eq-dust} \psi'' + 24Ea\psi = 0.
\end{equation}
The Bessel functions are solutions of the above equation. Therefore, the time dependent solutions
are as follows
\begin{equation} \label{bessel} \Psi_E =
e^{-iEt}\sqrt{a}\biggr[c_1J_{\frac{1}{3}}\biggr(\frac{\sqrt{96E}}{3}a^{\frac{3}{2}}\biggl) +
c_2Y_{\frac{1}{3}}\biggr(\frac{\sqrt{96E}}{3}a^{\frac{3}{2}}\biggl)\biggl] .
\end{equation}
Now, the wave packets can be constructed, by superposing these eigenfunctions with the following structure
\begin{equation}
\Psi(a,t) = \int_0^\infty A(E)\Psi_E(a,t)dE  .
\end{equation}
We choose $c_2 = 0$, for satisfying the first boundary condition (\ref{boundary}). By choosing
$A(E)$ as a quasi-gaussian weight factor and defining $r = \frac{\sqrt{96E}}{3}$, analytical
expressions for the wavepacket can be found
\begin{equation}
\Psi(a,t) = \sqrt{a}\int_0^\infty r^{\nu + 1}e^{-\gamma r^2 + i\frac{3}{32}r^{2}
t}J_\nu(ra^\frac{3}{2})dr,
\end{equation}
where $\nu = \frac{1}{3}$ and $\gamma$ is an arbitrary positive constant. The above integral is known \cite{gradshteyn}, and the wave packet takes
the form
\begin{equation}
\label{wp} \Psi(a,t) = a\frac{e^{-\frac{a^{3}}{4B}}}{(-2B)^{\frac{4}{3}}},
\end{equation}
where $B = \gamma - i\frac{3}{32}t$.  Following the many worlds interpretation of quantum mechanics \cite{everett}, we may write the expected value
for the scale factor $a$ as
\begin{equation}
<a>(t) = \frac{\int_0^\infty a\Psi(a,t)^*a\Psi(a,t)da} {\int_0^\infty a\Psi(a,t)^*\Psi(a,t)da} .
\end{equation}
which yields
\begin{equation} <a>(t) \propto
\biggr[\frac{9}{(32)^2\gamma^2}t^2 + 1\biggl]^\frac{1}{3} .
\end{equation}
These solutions represent a bouncing Universe, with no singularity, which goes asymptotically to
the corresponding flat classical models for late times (Fig. 1)
\begin{equation}
a(t) \propto t^{2/3}.
\end{equation}
\begin{figure}
\centering
  \includegraphics[width=8cm]{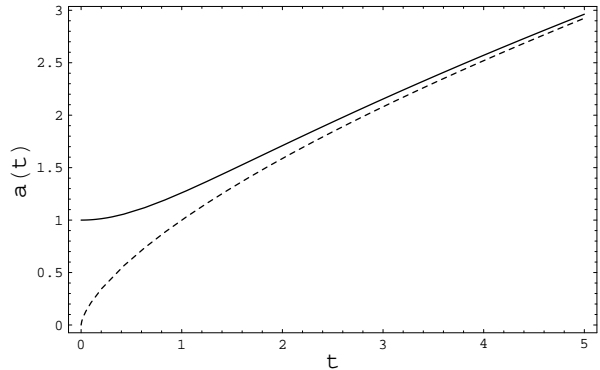}\\
  \caption{The behavior of the expected value
for the scale factor $\langle a\rangle(t)$ (solid line) and the
classical scale factor $a(t)$ (dashed line).}
\end{figure}

In the case $k=1$ the time-independent Wheeler-DeWitt equation (\ref{sle2}) reduces to
\begin{equation}
-{\psi}^{\prime \prime}(a) + \left(- 24Ea + 144a^{2}\right){\psi}(a)=0.
\end{equation}
Defining new variable $x=12a - E$ we find
\begin{equation}\label{k1}
-\frac{d^{2}\psi}{dx^{2}}+\left[- \frac{E^{2}}{144}+\frac{x^{2}}{144} \right]\psi(a) =0.
\end{equation}
Equation (\ref{k1}) is formally identical to the time-independent Schr\"odinger equation for a
harmonic oscillator with unit mass and energy $\lambda$
\begin{equation}
-\frac{d^{2}\psi}{dx^{2}}+\left[- 2\lambda+w^{2}x^{2}\right]\psi(x) =0,
\end{equation}
where $2\lambda = E^{2}/144$ and $w=1/12$. As much as the allowed
values of $\lambda$ are $n+1/2$, the possible values of $E$ are
\begin{equation}
E_{n}=\sqrt{12(2n+1)}\,\, , \mbox{\hspace{0.8cm}} n=0,1,2,...\quad .
\end{equation}
Thus, the  stationary solutions are
\begin{equation}
{\Psi}_{n}(a,t)=e^{-iE_{n}t}{\varphi}_{n}\left(12a - E_{n}\right), \label{k1-final}
\end{equation}
where
\begin{equation}
{\varphi}_{n}(x)=H_n\bigg(\frac{x}{\sqrt{12}}\bigg)e^{-x^2/24}\,\, ,
\label{dust7}
\end{equation}
with $H_n$  the $n$-th Hermite polynomial. The wave functions (\ref{k1-final}) are similar to the
stationary quantum wormholes as defined in \cite{Hawking}. However, neither of the boundary
conditions (\ref{boundary}) can be satisfied by the these wave functions.

In the $k=-1$ case the equation (\ref{sle2}) reduces to
\begin{equation}
{\psi}^{\prime \prime}(a) + \left(24Ea + 144a^{2}\right){\psi}(a)=0.
\end{equation}
where the solutions are
\begin{equation}
\label{whittaker} \Psi (a,t)=e^{-iEt}(12a+E)^{-1/2}\left\{ C_{1}M_{\frac{iE^2}{48},\frac{1}{4}}(\frac{i(12a+E)^2}{12}) +
C_{2}W_{\frac{iE^2}{48},\frac{1}{4}}(\frac{i(12a+E)^2}{12})\right\}
\end{equation}
where $M_{\kappa , \lambda}$ and $W_{\kappa , \lambda}$ are Whittaker functions. The Whittaker
functions do not automatically vanish at $a=0$. Therefore, in order to satisfy $\Psi(0,t)=0$ it is
necessary to take both $C_{1}\neq 0$ and $C_{2}\neq 0$, the same is applied to the second of the
boundary conditions (\ref{boundary}).

\section{Conclusions}
In this work we have investigated closed, flat, and open
minisuperspace FRW quantum cosmological models ($k=1,0,-1$) with
chaplygin gas Universes. The use of Schutz's formalism for chaplygin
gas allowed us to obtain a SWD equation in which the only remaining
matter degree of freedom played the role of time. We have obtained
eigenfunctions and therefore acceptable wave packets were
constructed by appropriate linear combination of these
eigenfunctions. The time evolution of the expectation value of the
scale factor has been determined in the spirit of the many-worlds
interpretation of quantum cosmology.
\section*{acknowledgements}
The authors thank H. R. Sepangi for useful discussions and comments.

\end{document}